# First-Principles Insights into Excitonic and Electron-Phonon Effects in van der Waals Heterostructures

Mohammad Ali Mohebpour[1], Carmine Autieri[2], Meysam Bagheri Tagani[1,*]

*Coresponding author: m_bagheri@guilan.ac.ir

[1]*Department of Physics, University of Guilan, P. O. Box 41335-1914, Rasht, Iran*

[2]*International Research Centre Magtop, Institute of Physics, Polish Academy of Sciences, Aleja Lotników 32/46, 02668 Warsaw, Poland*

## Abstract

Motivated by the successful synthesis of isolated $ZrS_2$ and $HfS_2$ transition metal dichalcogenide (TMD) monolayers and inspired by their nearly identical lattice constants, we construct and investigate a vertical $ZrS_2/HfS_2$ van der Waals (vdW) heterostructure. Using first-principles calculations based on density functional theory (DFT) and many-body perturbation theory (MBPT), we explore its electronic, optical, and excitonic properties, with particular emphasis on excitonic effects and their temperature dependence. Based on the GW method, the $ZrS_2/HfS_2$ vdW heterostructure exhibits an indirect band gap of 2.60 eV with a Type-I band alignment. The optical gap of the heterostructure is found to be 2.64 eV, with an exciton binding energy of 0.71 eV, both reduced compared to those in the isolated monolayers. Moreover, we investigate the temperature-dependent optoelectronic behavior of the heterostructure, considering electron-phonon coupling. A zero-point renormalization of 0.04 eV in the direct band gap is observed. While the direct band gap decreases monotonically with temperature from 0 K to 400 K, the indirect band gap displays a non-monotonic trend. As a result, the absorption spectrum undergoes a meaningful redshift with increasing temperature. At room temperature, the optical gap of the heterostructure is reduced to 2.51 eV and the exciton binding energy to 0.63 eV. Our findings highlight the important role of electron-phonon interaction in the optoelectronic response of $ZrS_2/HfS_2$ vdW heterostructure, supporting its use in high-performance optoelectronic devices.

## I. INTRODUCTION

Two-dimensional (2D) semiconducting transition metal dichalcogenides (TMDs) have emerged as a versatile class of materials with remarkable electronic, optical, and excitonic properties, making them attractive for advanced optoelectronic devices [1-5]. By vertically stacking two different TMD monolayers, TMD heterostructures are formed, unlocking novel physical phenomena not present in the individual monolayers [6-11]. These heterostructures provide unique features such as tunable band alignment [12, 13], long-lived interlayer excitons [14, 15], tunable optical and excitonic responses [16, 17], broadband optical absorption or emission [17, 18], and twist-angle-dependent features like moiré excitons [19-22], which are favorable for optoelectronic applications. $MoS_2/WS_2$ and $MoSe_2/WSe_2$ are among the most widely studied TMDs due to their Type-II band alignment, which enables the creation of long-lived interlayer excitons [6, 16]. Beyond Mo- and W-based TMD heterostructures, other combinations such as $ZrS_2/HfS_2$ [23-25] show great promise for optoelectronic applications due to their strong light-matter interactions. Recently, two-dimensional heterojunctions based on $HfSe_2$ were synthesized [26] and it was proved to be a candidate for High-Performance phototransistors. The flexibility in material choice and stacking configurations allows for tunable properties, turning the TMD heterostructures into promising candidates for applications in solar cells [27], light-emitting diodes (LED) [28], photodetectors [29], and lasers [30]. As a result, understanding the fundamental properties of these van der Waals (vdW) heterostructures, particularly through first-principles calculations, is essential for designing next-generation optoelectronic devices and unlocking their potential in practical applications.

Electron-hole pairs (excitons) play an important role in the optical properties of 2D materials due to reduced dielectric screening and strong quantum confinement [31-33]. In TMD heterostructures like $MoS_2/WS_2$ and $MoSe_2/WSe_2$, excitons dominate the optical response. Therefore, accounting for electron-hole interactions is crucial for accurate optical modeling. Additionally, electron-phonon coupling significantly influences electronic band structure and optical absorption, making the optoelectronic properties of TMD heterostructures highly sensitive to temperature variations [34, 35]. For example, Lie et al. [36] investigated the temperature-dependent optical constants (T=4.5 - 500 K) of $MoS_2$, $MoSe_2$, $WS_2$, and $WSe_2$ monolayers. They observed that the optical band gaps of the monolayers decrease with increasing temperature, due to enhanced electron-phonon coupling. They also reported a noticeable redshift in the absorption spectra of the monolayers as temperature increases. These findings were confirmed by the subsequent study conducted by Bezerra et al. [37]. In a separate study, Sharma et al. [38] explored the influence of electron-phonon coupling on the electronic properties of the $ZrS_2$ monolayer, revealing that this interaction significantly alters its characteristics. They found that band structure renormalization induced by electron-phonon coupling not only narrows the band gap but also modifies the band dispersion, which strongly affects the optical properties of the monolayer. Similarly, another study [39] reported a zero-point renormalization (ZPR) of 0.03 eV in the band gap of the $HfS_2$ monolayer, which is attributed to strong electron-phonon coupling. Additionally, numerous studies [40-42] offer valuable insights into the temperature-dependent optoelectronic properties of TMD heterostructures, emphasizing that electron-phonon coupling is crucial for achieving agreement between experimental observations and theoretical calculations. Overall, one can conclude that accurately capturing the optoelectronic properties of TMD heterostructures requires careful consideration of electron-hole interactions and electron-phonon coupling.

In this work, we focus on two less-studied TMD monolayers, 1T-ZrS$_2$ and 1T-HfS$_2$, which exhibit significant promise both as individual monolayers and as layers of a vdW heterostructure. Monolayer 1T-ZrS$_2$ has been experimentally synthesized using the low-pressure chemical vapor deposition [43]. Monolayer 1T-HfS$_2$ has been synthesized through mechanical exfoliation [44]. The lattice constants of these TMD monolayers, as documented in the literature [24], indicate the potential for creating a stable vdW heterostructure with a minimal lattice mismatch. Accordingly, using density functional theory (DFT) and many-body perturbation theory (MBPT), we explore the electronic, optical, and excitonic properties of ZrS$_2$/HfS$_2$ vdW heterostructure, with special focus on its excitonic features and temperature-dependent behaviors arising from electron-phonon coupling. As far as we are aware, no prior study has investigated the optoelectronic properties of the ZrS$_2$/HfS$_2$ vdW heterostructure while accounting for both electron-hole and electron-phonon interactions, highlighting the novelty of our work. In addition, former studies [23, 24] have reported conflicting results about the type of band alignment in this heterostructure. Therefore, a comprehensive investigation that resolves these discrepancies and explores new aspects of the heterostructure is needed. We further analyze the electronic and optical properties of the individual ZrS$_2$ and HfS$_2$ monolayers, to provide a reliable basis for comparison with the heterostructure and to validate the accuracy of our computational approach through consistency with previous studies. Our findings underscore the significant role of electron-phonon interaction in the optoelectronic response of ZrS$_2$/HfS$_2$ vdW heterostructure, supporting its potential for high-performance optoelectronic applications.

## II. COMPUTATIONAL METHODS

Ground-state DFT calculations were performed with the Quantum-ESPRESSO package [45, 46]. The Perdew–Burke–Ernzerhof (PBE) functional within the framework of the generalized gradient approximation (GGA) described the exchange-correlation potentials [47]. The optimized Vanderbilt norm-conserving (NC) pseudopotentials provided by PseudoDojo [48, 49] represented the valence electrons. The DFT-D3 method accounted for vdW correction [50]. A kinetic energy cutoff of 50 Ry ensured accurate electronic structure calculations. A $16 \times 16 \times 1$ Monkhorst-Pack k-point mesh sampled the Brillouin zone. A vacuum spacing of 20 Å for monolayers and 25 Å for the heterostructure prevented interactions in the non-periodic direction.

Quasiparticle (QP) energies were computed using the single-shot GW approximation (G$_0$W$_0$), in which the screened interaction was modeled via the plasmon-pole approximation (PPA) [51]. According to convergence tests, a $16 \times 16 \times 1$ k-point mesh was used in the Brillouin zone sampling. Additionally, an energy cutoff of 50 Ry was applied to the exchange component, while a 5 Ry cutoff was used for the correlation component. Furthermore, 60 and 120 empty bands were included in the GW calculations for the monolayers and the heterostructure, respectively, to guarantee complete convergence. The excitonic absorption spectra were determined by solving the Bethe-Salpeter equation (BSE) [52]. Ten valence bands (VBs) and ten conduction bands (CBs) were included in the BSE calculation. All the GW and BSE calculations were performed using the YAMBO code [53, 54]. The results of convergence tests for the k-point mesh, energy cutoffs, and number of bands are provided in Figures S1–S4 and discussed in Text S1 of the Supplemental Material. In addition to optical calculations at the BSE level, which includes excitonic effects, the absorption spectra were determined using the independent-particle approximation (DFT-IPA), which excludes these interactions, to see the impact of excitons on optical properties.

Many-body perturbation theory (MBPT) is used to describe temperature-dependent electronic states. In this approach, electron-phonon interactions are treated by expanding the nuclear displacement to first and second order, known as the Fan and Debye-Waller (DW) terms. The resulting interacting Green's function can be written as:

$$G_{nk}(\omega, T) = [\omega - \epsilon_{nk} - \Sigma_{nk}^{Fan}(\omega, T) - \Sigma_{nk}^{DW}(T)]^{-1}$$

where $\epsilon_{nk}$ is the Kohn-Sham ground-state eigenvalues for frozen atoms. $\Sigma^{Fan}$ is the Fan contribution

$$\Sigma_{nk}^{Fan}(i\omega, T) = \Sigma_{n'q\lambda} \frac{|g_{nn'k}^{q\lambda}|^2}{N} \left[ \frac{N_{q\lambda}(T) + 1 - f_{n'k-q}}{i\omega - \epsilon_{n'k-q} - \omega_{q\lambda}} + \frac{N_{q\lambda}(T) + f_{n'k-q}}{i\omega - \epsilon_{n'k-q} + \omega_{q\lambda}} \right],$$

and $\Sigma^{DW}$ is the Debye-Waller term

$$\Sigma_{nk}^{DW}(T) = -\frac{1}{2} \Sigma_{n'q\lambda} \frac{\Lambda_{nn'k}^{q\lambda}}{N} \left[ \frac{2N_{q\lambda}(T) + 1}{\epsilon_{nk} - \epsilon_{n'k}} \right].$$

Here, $N_{q\lambda}$ and $f_{n'k-q}$ are the Bose-Einstein and Fermi-Dirac distribution functions, while $N$ is the number of $q$ points in the Brillouin zone. For more details on the electron-phonon matrix elements, see Ref [55].

## III. RESULTS AND DISCUSSION

### A. ZrS$_2$ and HfS$_2$ monolayers

The crystal structure of the isolated ZrS$_2$ and HfS$_2$ monolayers is first investigated, as shown in Figure S5. Both ZrS$_2$ and HfS$_2$ monolayers exhibit 1T phase that belongs to the P-3m1 space group no. 164. The relaxed lattice constant is 3.68 Å for $ZrS_2$ monolayer and 3.65 Å for $HfS_2$. Also, the bond length is 2.57 Å for $ZrS_2$ monolayer and 2.55 Å for $HfS_2$. Despite the slight differences in lattice constants and bond lengths resulting from different atomic sizes, $ZrS_2$ and $HfS_2$ monolayers maintain the same thickness of 2.89 Å. Our results are consistent with previous studies [26, 56-58].

In Figure 1(a), we compare the PBE and GW band structures of $ZrS_2$ monolayer, revealing that it is an indirect band gap semiconductor with the valence band maximum (VBM) located at the Γ point and the conduction band minimum (CBM) at the M point. The PBE band gap is 1.19 eV, while the GW band gap is 2.97 eV. The PBE value is in close agreement with the previously reported 1.20 eV [59, 60]. Also, the GW gap is consistent with 2.81 eV reported in the literature [25]. The calculated PBE band gap is smaller than the GW band gap due to the lack of consideration for electron-electron interactions and self-energy corrections.

Next, we analyze the absorption spectrum of $ZrS_2$ monolayer as shown in Figure 1(b). The spectrum was calculated at two levels of theory: one without electron-hole interactions (using the simple IPA method) and one with these interactions included (using the BSE method). The IPA curve (blue) shows the absorption pattern when excitonic effects are not considered. This curve shows less prominent peaks and weaker absorption. In this case, the absorption edge appears near 1.50 eV. The gradual increase in absorption, rather than sharp peaks, underscores the absence of electron-hole interactions. On the contrary, the BSE curve (red) captures the influence of excitonic effects. This curve exhibits a strong, distinct peak near 3 eV, coming from the transition between the highest VB and the lowest CB at the M point. At this level, the absorption spectrum is significantly blue-shifted compared to that of the IPA level. The first absorption peak in this case is at 2.75 eV, contributed by the vertical transition between the VBM and the lowest CB at the Γ point. This peak corresponds to an exciton with a binding energy of 0.95 eV, which is very stable. Figure 1(b) also shows that the first absorption peak has considerable oscillator strength, confirming it as a bright exciton. The inset of Figure 1(b) illustrates the oscillator strength of the six lowest-energy excitons, showing that the $2^{nd}$, $3^{rd}$, and $5^{th}$ excitons are bright and optically active, with the $2^{nd}$ and $3^{rd}$ being degenerate. This means that the $1^{st}$, $4^{th}$, and $6^{th}$ excitons are dark and optically inactive. The $2^{nd}$ exciton is correlated with the first absorption peak. In other words, there is an optically forbidden dark exciton below 2.75 eV (first absorption peak), stemming from the direct transitions between the highest VB and lowest CB along the Γ-M path.

Figure 1(c) displays bright and dark excitons within the 2.6 to 3.4 eV range, demonstrating how these excitonic states give rise to a rich and tunable optical response within $ZrS_2$ monolayer. Figure 1(d) shows the amplitude of the first bright exciton, which reveals a sharp dominant peak, an indicator of a highly localized exciton. The narrow linewidth of this peak reflects an excitonic state with minimal dispersion, a characteristic that benefits optoelectronic applications.

Moreover, Figure 1(e) reveals that the wave function of the first bright exciton is highly localized around the center, suggesting that the exciton has a strong binding energy and a small Bohr radius. The symmetry and spatial distribution of the wave function indicate that the exciton is confined within a limited space, which is consistent with the amplitude observed for this exciton.

All the calculations discussed so far focus on $ZrS_2$ monolayer. Next, we delve into the electronic and optical properties of $HfS_2$ monolayer. From Figure 2(a), one can find that $HfS_2$ monolayer is also an indirect band gap semiconductor with the VBM located at the Γ point and the CBM at the M point. The PBE band gap is 1.34 eV, while the GW band gap is 3.07 eV. The PBE value closely matches the previously reported values of 1.36 eV [25] and 1.31 eV [61]. The GW gap aligns with 2.94 eV reported in Ref. [62]. The calculated band gaps for $HfS_2$ monolayer are larger than those for $ZrS_2$ monolayer, however, the overall shape of their band structures remains similar.

The absorption spectrum of $HfS_2$ monolayer is plotted in Figure 2(b). As it is clear, both the IPA and BSE absorption spectra display two distinct peaks. The BSE curve is remarkably blue-shifted compared to the IPA curve because of the synergic effects of electron-electron and electron-hole interactions. At the IPA level, the absorption is relatively weak and begins around 2 eV. While the first absorption peak in the BSE spectrum appears at 3.30 eV and arises from a vertical transition between the VBM and the lowest CB at the Γ point. The second absorption peak, observed at 3.45 eV, corresponds to an exciton stemming from a vertical transition between the highest VB and lowest CB at the M point. More importantly, there is a slight bump at 3.20 eV, which is attributed to a bright exciton arising from transitions between the same bands along the Γ–M path in the Brillouin zone. The binding energy of this exciton is 0.99 eV. Our BSE calculations indicate that the optical gap and the exciton binding energy of $HfS_2$ monolayer are larger than those of $ZrS_2$ monolayer. This suggests that a vdW heterostructure combining these two monolayers could efficiently absorb light over a wider range of energy. Figure 2(b) also shows that the two main peaks of absorption spectrum correspond to bright excitons due to their high oscillator strengths. In particular, the 1$^{st}$ exciton, correlated with the slight bump at 3.20 eV, shows an oscillator strength that confirms its bright nature. The inset in Figure 2(b) presents the oscillator strengths of the seven lowest-energy excitons, indicating that only the 1$^{st}$ and 4$^{th}$ excitons are bright and optically active, while the others (2$^{nd}$, 3$^{rd}$, 5$^{th}$, 6$^{th}$, and 7$^{th}$) remain dark.

From Figure 2(c), one can identify four distinct bright excitonic states (highlighted by red lines), indicating that $HfS_2$ monolayer slightly absorbs and emits light within this energy range. Particularly, there is no optically forbidden dark exciton before the first bright exciton at 3.20 eV. This allows us to directly attribute the onset of optical absorption to an optically active transition. However, as shown in Figure 3(d), the amplitude of the first exciton has multiple peaks, indicating that its state is more complex than a single resonance. The wave function of this exciton is highly localized and lacks symmetry (see Figure 2(e)), which aligns with the calculated amplitude.

### B. $ZrS_2/HfS_2$ vdW heterostructure

Following our investigation of the electronic and optical properties of $ZrS_2$ and $HfS_2$ monolayers, results fully consistent with earlier findings, we now shift our focus to $ZrS_2/HfS_2$ vdW heterostructure. The nearly identical lattice constants of the individual $ZrS_2$ and $HfS_2$ monolayers result in a minimal mismatch (0.5 %), which facilitates the formation of a stable vdW heterostructure, a feature that shows promise for future optoelectronic applications. Although $ZrS_2/HfS_2$ vdW heterostructure has not yet been synthesized experimentally, theoretical investigations have confirmed its thermal and dynamical stabilities [23, 24, 26]. Consistent with earlier studies [23, 24], our systematic exploration of six different stacking configurations (as shown in Figure S6 and detailed in Table S1) indicates that the AA stacking configuration is energetically more favorable than the other stacking configurations. Moreover, the electronic band structures of these stacking configurations show similar characteristics (as shown in Figure S7 and discussed in Text S2), leading us to focus exclusively on the AA stacking configuration. The resulting lattice constant of 3.68 Å and the interlayer distance of 2.90 Å for this stacking configuration agree with previous reports [23, 24, 26] supporting the validity of our results. The $ZrS_2/HfS_2$ heterostructure has different symmetries with respect to the monolayers giving the chance to manipulate the properties of the excitons.

The electronic properties of $ZrS_2/HfS_2$ vdW heterostructure reflect the intrinsic characteristics of the individual $ZrS_2$ and $HfS_2$ monolayers and the modifications induced by interlayer interactions. As shown in Figure 3(a), the heterostructure preserves the indirect band gap, with the VBM located at the $\Gamma$ point and the CBM at the M point, features consistent with the electronic structure of the isolated monolayers. However, the size of the band gap is noticeably reduced compared to the monolayers. Specifically, the DFT-PBE calculations yield an indirect band gap of 1.18 eV, which increases to 2.60 eV with the GW corrections; both values remain lower than those of the isolated monolayers. All the calculated band gaps are provided in Table II. Also, as shown in Figure 3(b), the spatial distribution of the wave functions at the band extrema reveals a Type-I band alignment; the VBM is shared between the two layers, while the CBM is predominantly localized on the $ZrS_2$ layer. This alignment promotes strong carrier confinement and facilitates efficient electron-hole recombination, making $ZrS_2/HfS_2$ vdW heterostructure a promising candidate for high-performance optoelectronic devices such as light-emitting diodes and lasers.

Figure 3(c) shows the optical absorption spectrum of ZrS$_2$/HfS$_2$ vdW heterostructure calculated using the DFT-IPA (blue curve) and the GW-BSE (red curve) methods. At the IPA level, the absorption starts at around 2.00 eV, displaying a shoulder around 2.15 eV and a double-hump structure between 2.20 and 2.60 eV, underscoring the absence of electron-hole interactions. However, when excitonic effects are included using the BSE formalism, the absorption spectrum undergoes noticeable changes; a pronounced excitonic peak emerges at 2.68 eV, exhibiting a notable blueshift compared to the IPA onset, followed by intense peaks at 3.02 eV and 3.28 eV. The first absorption peak corresponds to the degenerate 4$^{th}$ and 5$^{th}$ excitons, meaning three lower-energy excitons (1$^{st}$, 2$^{nd}$, and 3$^{rd}$) lie below this threshold. As shown in the inset of Figure 3(c), the 2$^{nd}$, 4$^{th}$, and 5$^{th}$ excitons are bright and optically active, while the 1$^{st}$ and 3$^{rd}$ excitons are dark and optically inactive. The 2$^{nd}$ exciton, as the first bright exciton, appearing at 2.64 eV, results from direct transitions between the VBM and the lowest CB at the Γ point, with a binding energy of 0.71 eV. The second and third absorption peaks, at 3.02 eV and 3.28 eV, respectively, also correspond to bright excitons due to their considerable oscillator strengths. Unlike the isolated ZrS$_2$ and HfS$_2$ monolayers, the BSE spectrum of ZrS$_2$/HfS$_2$ vdW heterostructure is characterized by three prominent peaks. The first and second peaks dominantly arise from the ZrS$_2$ layer, while the third peak originates from the HfS$_2$ layer. Overall, the heterostructure exhibits an optical gap of 2.64 eV and an exciton binding energy of 0.71 eV -both lower than those in the isolated monolayers- but with stronger absorption and broader linewidth, supporting its potential for advanced optoelectronic applications.

Figure 4(a) shows the exciton spectrum of ZrS$_2$/HfS$_2$ vdW heterostructure in the energy range of 2.6 to 3.4 eV, highlighting the presence of bright and dark excitonic states. This spectrum reveals a high density of bright excitonic states within this range, much higher than the monolayers, indicating strong light-matter interaction. These bright excitons correspond to transitions allowed by optical selection rules, making them directly observable in photoluminescence experiments. Figure 4(b) shows the amplitudes of the first six excitons. The 2$^{nd}$ exciton -the first bright one- has a broader, more complex profile, indicating contributions from several transitions and a more delocalized nature in reciprocal space. On the contrary, the 4$^{th}$ exciton -the second bright one- shows a sharp, narrow peak, suggesting contributions from a single transition and a more localized character in reciprocal space. Figures 4(c) and 4(d) illustrate the real-space wave functions of the 2$^{nd}$ and 4$^{th}$ excitons, respectively. Both excitons are intralayer, meaning that the electrons and holes are confined within the same layer, specifically, the ZrS$_2$ layer of the heterostructure. The 2$^{nd}$ exciton is more localized, with the electron density tightly concentrated around the hole, indicating strong electron-hole binding energy and a small exciton radius. Conversely, the 4$^{th}$ exciton is more delocalized, extending the electron density over a larger region. In summary, ZrS$_2$/HfS$_2$ vdW heterostructure supports a rich excitonic structure, with a high density of bright excitons in the visible energy range and a mixture of localized and delocalized intralayer excitonic states, underscoring the potential of the heterostructure for optoelectronic applications.

Electron-phonon coupling is crucial in determining the electronic and optical properties of vdW heterostructures. As lattice vibrations interact with charge carriers, they can lead to band gap renormalization and a redshift in the absorption spectrum with increasing temperature. Therefore, accurately accounting for these effects is essential for understanding and predicting the temperature-dependent behavior of the heterostructure in optoelectronic applications. Figure 5 clearly illustrates how the electronic band structure of $ZrS_2$/$HfS_2$ vdW heterostructure evolves with temperature due to electron-phonon coupling. As the temperature increases from 50 K to 400 K, this interaction causes noticeable changes in the valence and conduction band states. At lower temperatures (50 K – 100 K), the band extrema remain unchanged, with the VBM and CBM located at the $\Gamma$ and M points, respectively, similar to the configuration observed in the absence of electron-phonon coupling, shown in Figure 3(a). However, from 150 K, the VBM begins to deviate from the $\Gamma$ point. By 350 K, both the VBM and CBM move closer together in momentum space, before separating again at 400 K. This behavior reflects a non-monotonic renormalization of the band gap.

Figure 6(a) presents the temperature-dependent variation of the direct and indirect band gaps of $ZrS_2$/$HfS_2$ vdW heterostructure, over the range of 0 - 400 K. The direct band gap shows a monotonic decrease with increasing temperature, dropping from 1.64 eV at 0 K to around 1.54 eV at 400 K. This behavior reflects band gap renormalization caused by stronger electron-phonon interactions at higher temperatures. A similar behavior has been observed in the isolated $ZrS_2$ [38] and $HfS_2$ [39] monolayers. Without electron-phonon coupling, the direct band gap of the heterostructure is 1.68 eV, while including this interaction at 0 K reduces it to 1.64 eV. This small difference of 0.04 eV is known as ZPR, arises from the lattice vibrations even at absolute zero.

In contrast, the indirect band gap shows a non-monotonic trend. It increases sharply from 1.18 eV at 0 K to a peak of 1.46 eV around 200 K, then gradually decreases to 1.38 eV at 400 K. The initial increase in the indirect band gap is likely caused by thermal expansion of the lattice and moderate phonon activity. Beyond 200 K, stronger phonon scattering leads to a shift in the band edges that reduces the band gap. Despite all the changes in band edges, $ZrS_2$/$HfS_2$ vdW heterostructure maintains its indirect band gap character and Type-I band alignment during the entire temperature range.

Besides the electronic structure, the absorption spectrum of $ZrS_2$/$HfS_2$ vdW heterostructure shows temperature-dependent behavior. As illustrated in Figure 6(b), each spectrum displays multiple peaks, in contrast to the simpler curve observed without electron-phonon coupling (Figure 3(c)). Although the overall shape of the absorption spectrum remains consistent across temperatures, the variations in the peaks intensity and position are apparent. As the temperature increases, the entire absorption spectrum exhibits a gradual redshift, particularly noticeable in the first peak at 2.5 eV. This trend is attributed to the temperature-induced narrowing of the direct band gap. In addition to the redshift, the peaks become broader and less intense with increasing temperature. Specifically, at 200 K, the features are more pronounced, with sharper and higher peaks. However, at 300 K and 400 K, the peaks become less intense. Remarkably, the high-energy region above 3.0 eV remains relatively stable, suggesting that higher electronic transitions are less influenced by phonon scattering. At 300 K, for example, the first bright exciton of the heterostructure appears at 2.51 eV with a binding energy of 0.63 eV, both lower than those in the absence of electron-phonon coupling, as provided in Table I.

Figure 6(c) shows the oscillator strength of excitons in $ZrS_2$/$HfS_2$ vdW heterostructure, considering electron-phonon coupling at 300 K. As can be seen, electron-phonon coupling increases the number of peaks with strong oscillator strength, which means there are more bright excitons at room temperature than at absolute zero. A higher density of bright excitons increases the absorption coefficient, improving the efficiency of light-harvesting devices. It can also enhance the intensity of light emission, which is valuable for designing advanced optoelectronic devices. Figure 6(d) indicates the real-space wave function of the $1^{st}$ bright exciton in $ZrS_2$/$HfS_2$ vdW heterostructure at 300 K, considering electron-phonon coupling. This wave function closely resembles the one observed without electron-phonon coupling, as shown in Figure 4(c). This similarity indicates that the spatial distribution of the $1^{st}$ bright exciton, specifically, its intralayer character remains robust, even when electron-phonon coupling is considered.

## IV. CONCLUSION

In summary, we have systematically investigated the electronic, optical, and excitonic properties of $ZrS_2/HfS_2$ van der Waals heterostructure within density functional theory, considering electron-phonon coupling and many-body perturbation theory. The electronic structure reveals that the heterostructure maintains an indirect band gap. The PBE- and GW-calculated band gaps are 1.18 eV and 2.60 eV, respectively, both lower than those of the isolated $ZrS_2$ and $HfS_2$ monolayers, indicating band gap narrowing upon the heterostructure construction. The Bethe-Salpeter equation calculations reveal that, unlike the isolated $ZrS_2$ and $HfS_2$ monolayers, the $ZrS_2/HfS_2$ heterostructure has a rich excitonic spectrum with three prominent absorption peaks. The results also show that the heterostructure has an optical gap of 2.64 eV and an exciton binding energy of 0.71 eV, both smaller than those of the isolated monolayers. The first and second bright excitons are located in the $ZrS_2$ monolayer. The effects of electron-phonon coupling on the electronic, optical, and excitonic properties were also examined in detail. The inclusion of this interaction induces a zero-point renormalization of 0.04 eV in the direct band gap. As temperature increases, the direct gap decreases monotonically, reaching 1.54 eV at 400 K. Meanwhile, the indirect gap exhibits a non-monotonic temperature dependence, increasing from 1.18 eV at 0 K to a maximum of 1.46 eV at 200 K before decreasing to 1.38 eV at 400 K. In the presence of electron-phonon coupling at 300 K, the optical gap of the heterostructure is reduced to 2.51 eV and the exciton binding energy to 0.63 eV. These changes underscore the significant role of electron-phonon interaction in the optoelectronic response of $ZrS_2/HfS_2$ heterostructure. Overall, our findings provide deep insight into the electronic and optical properties of this heterostructure and support its potential in next-generation optoelectronic devices.

ACKNOWLEDGMENTS

M. A. M acknowledges the funding support by Iran National Science Foundation (INSF) under project No.4023062. C.A. was supported by the "MagTop" project (FENG.02.01-IP.05-0028/23) carried out within the "International Research Agendas" programme of the Foundation for Polish Science, co-financed by the European Union under the European Funds for Smart Economy 2021-2027 (FENG).

CONFLICT OF INTEREST

The authors have no conflicts to disclose.

DATA AVAILABILITY

The data that support the findings of this study are available from the corresponding author upon reasonable request.

Table I. Electronic band gaps (PBE and GW), optical gaps, and exciton binding energies for $ZrS_2$ monolayer, $HfS_2$ monolayer, and $ZrS_2/HfS_2$ vdW heterostructure. The type of band gap is shown. For the heterostructure, all parameters are also reported with electron-phonon coupling at 300 K.

| Structure | PBE band gap (eV) | GW band gap (eV) | Optical gap (eV) | Binding energy (eV) |
|---|---|---|---|---|
| $ZrS_2$ monolayer | 1.19 (Indirect) | 2.97 (Indirect) | 2.75 | 0.95 |
| $HfS_2$ monolayer | 1.34 (Indirect) | 3.07 (Indirect) | 3.20 | 0.99 |
| $ZrS_2/HfS_2$ | 1.18 (Indirect) | 2.60 (Indirect) | 2.64 | 0.71 |
| $ZrS_2/HfS_2$ (300 K) | 1.43 (Indirect) | 2.93 (Indirect) | 2.51 | 0.63 |

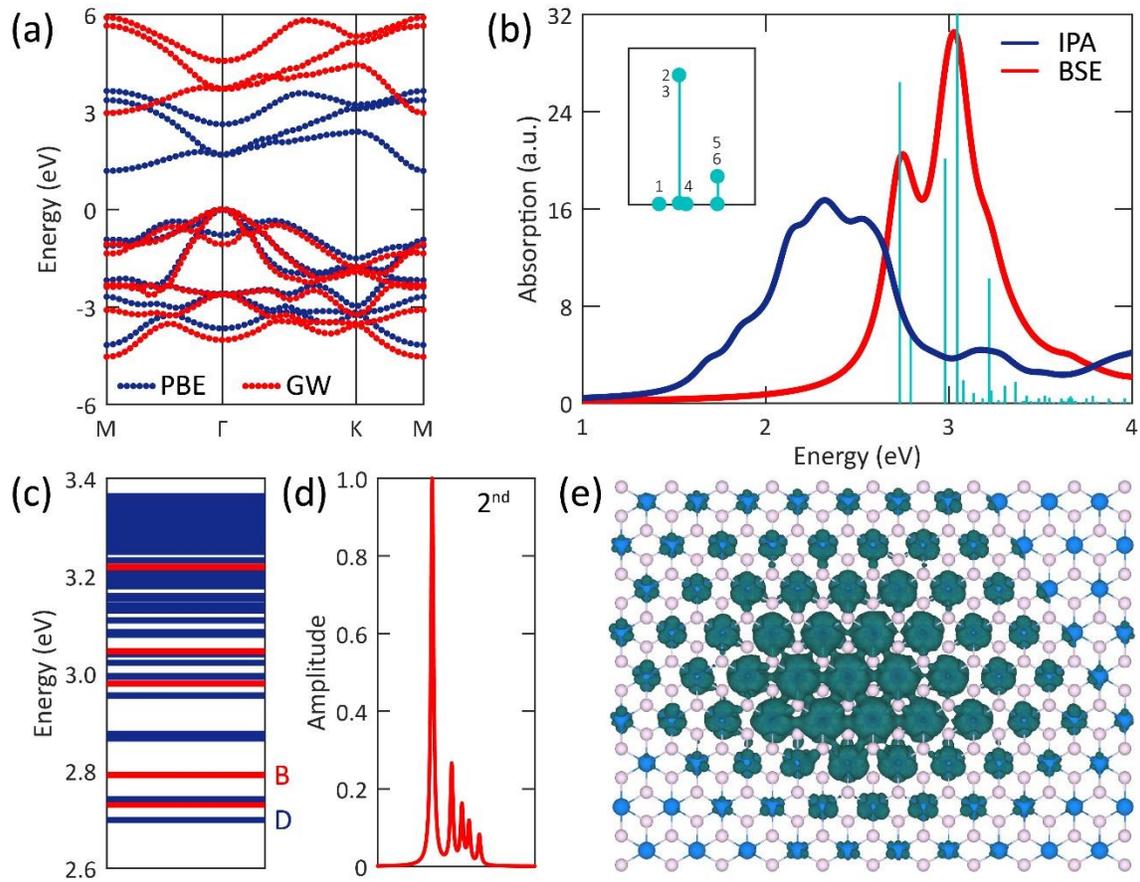

Figure 1. Electronic, optical, and excitonic properties of ZrS$_2$ monolayer. (a) Band structure of the monolayer at the PBE and GW approximations. (b) Absorption spectrum of the monolayer without (IPA) and with (BSE) electron-hole interactions. Cyan bars indicate the oscillator strength of optical transitions, normalized to one. The oscillator strength of the first six excitons is shown in the inset. (c) Exciton spectrum of the monolayer, showing the bright and dark excitons by red and blue lines, respectively. (d) Amplitude and (e) real-space wave function of the first bright exciton of the monolayer. Zr and S atoms are represented by blue and grey balls, respectively.

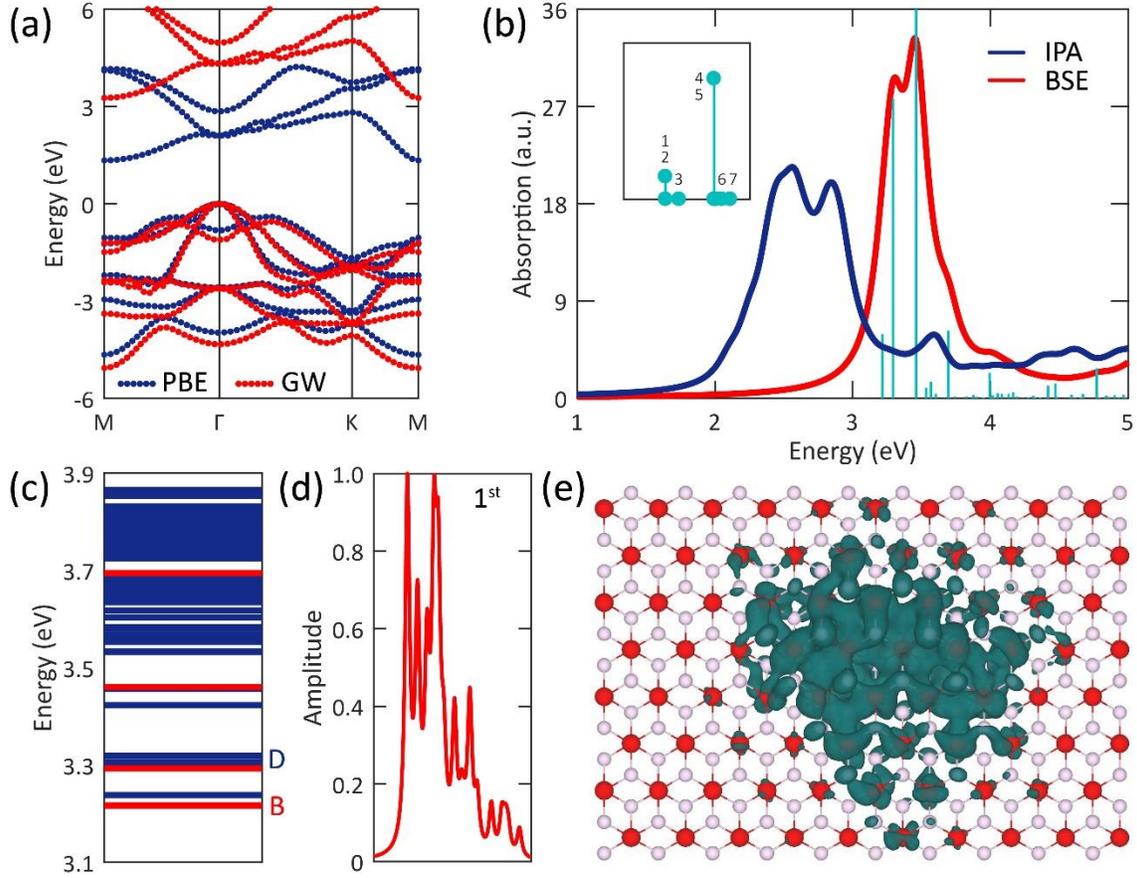

Figure 2. Electronic, optical, and excitonic properties of HfS$_2$ monolayer. (a) Band structure of the monolayer at the PBE and GW approximations. (b) Absorption spectrum of the monolayer without (IPA) and with (BSE) electron-hole interactions. Cyan bars indicate the oscillator strength of optical transitions, normalized to one. The oscillator strength of the first seven excitons is shown in the inset. (c) Exciton spectrum of the monolayer, showing the bright and dark excitons by red and blue lines, respectively. (d) Amplitude and (e) real-space wave function of the first bright exciton of the monolayer. Hf and S atoms are represented by red and grey balls, respectively.

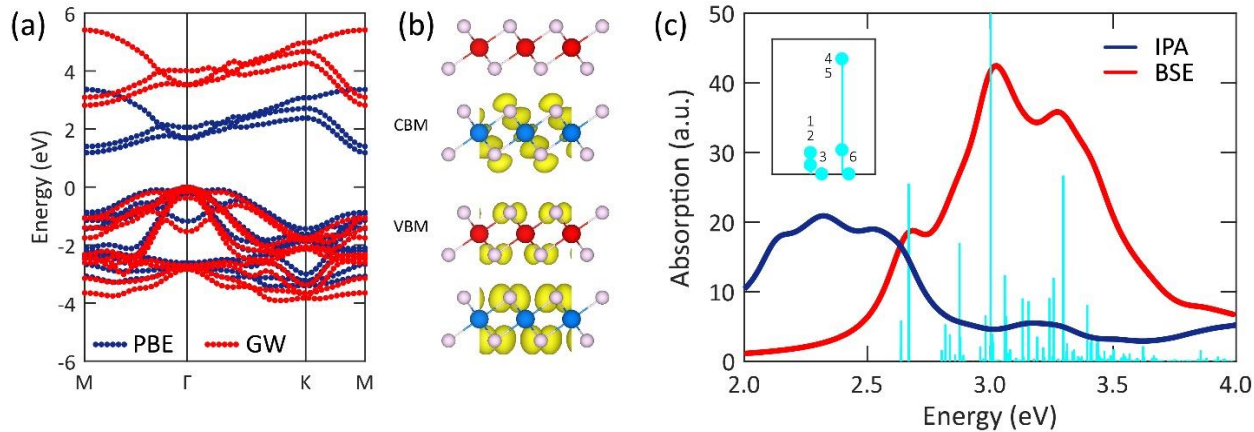

Figure 3. (a) Band structure of $ZrS_2/HfS_2$ vdW heterostructure at the PBE and GW approximations. (b) Electronic wave function at the CBM and VBM in the top and bottom panel, respectively. (c) Absorption spectrum of the heterostructure without (IPA) and with (BSE) electron-hole interactions. Cyan bars indicate the oscillator strength of optical transitions. The oscillator strength for the first six excitons is shown in the inset.

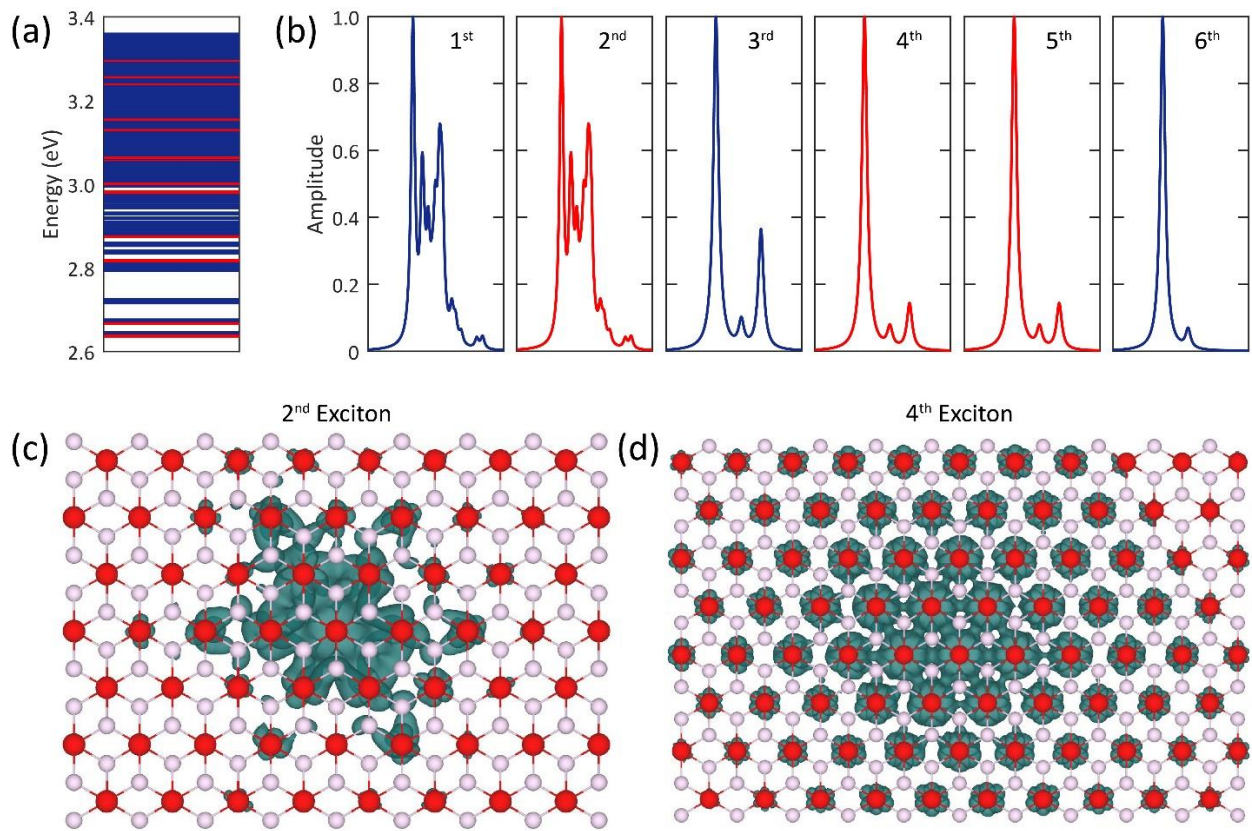

Figure 4. (a) Exciton spectrum of $ZrS_2/HfS_2$ vdW heterostructure, showing the bright and dark excitons by red and blue lines, respectively. (b) Amplitude of the six lowest-energy excitons. The first, third and sixth excitons are dark, the others are bright.

Real-space wave function of the first (c) and second (d) bright excitons of the heterostructure.

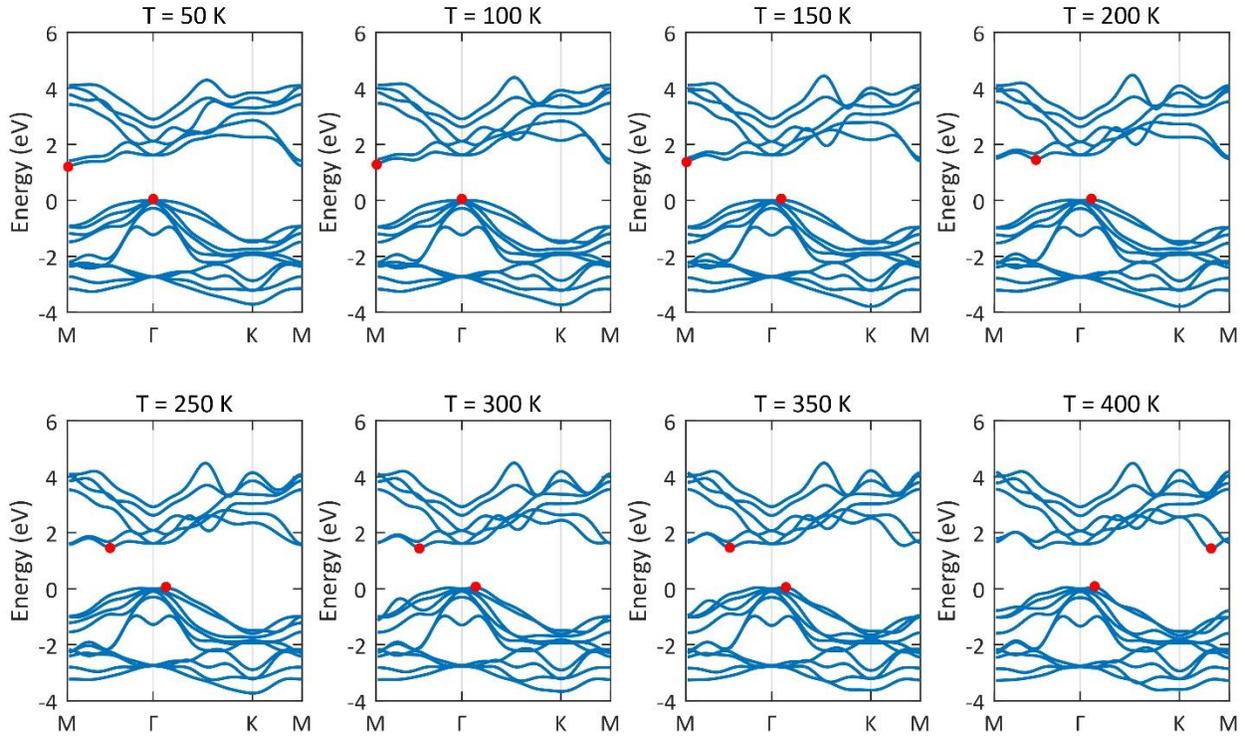

Figure 5. Temperature-dependent band structures of $ZrS_2$/$HfS_2$ vdW heterostructure, considering electron-phonon coupling at the PBE level. The location of the VBM and CBM are shown by red dots.

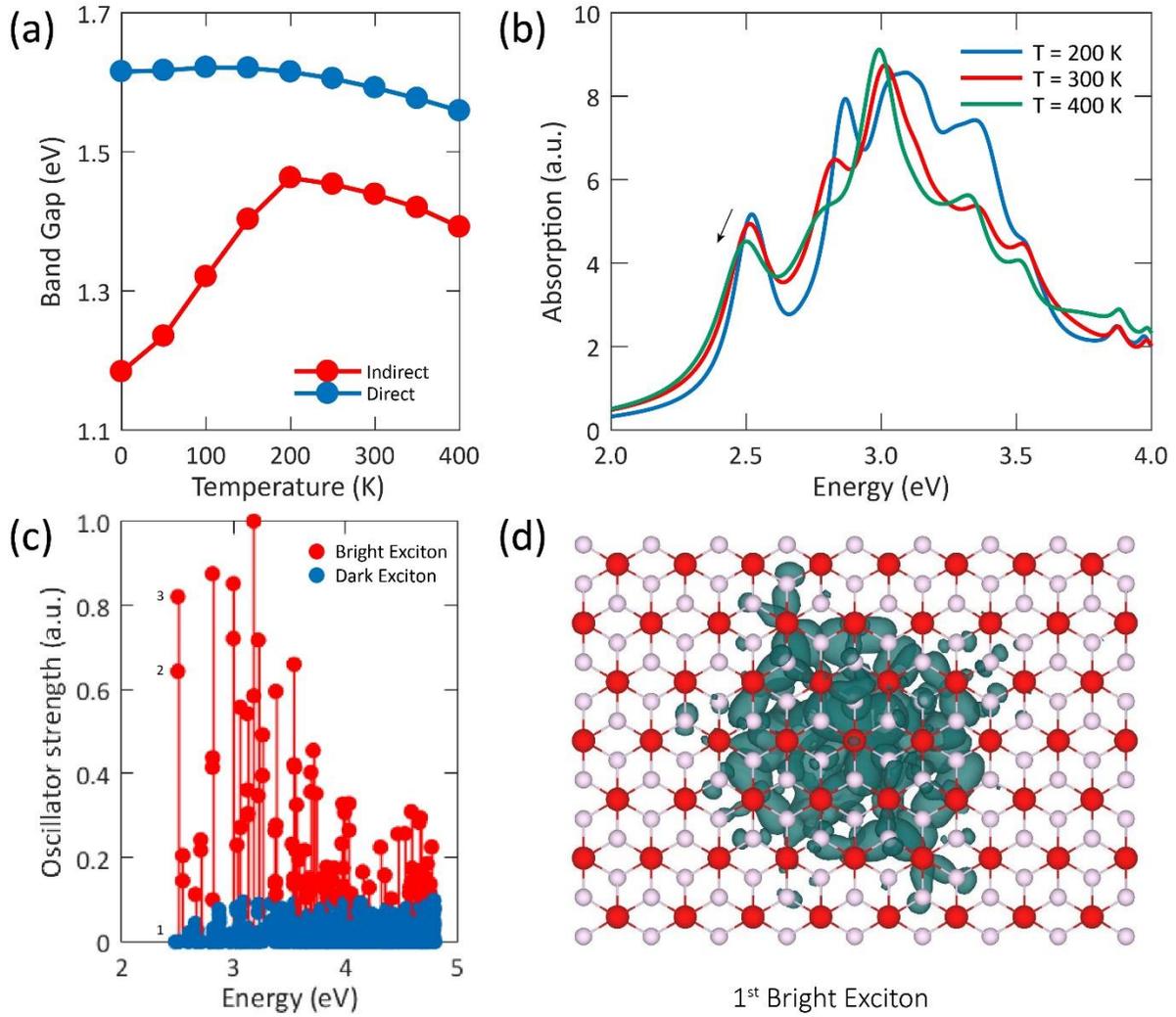

Figure 6. (a) Temperature-dependent band gaps of ZrS$_2$/HfS$_2$ vdW heterostructure, considering electron-phonon coupling at the PBE level. (b) Absorption spectrum of the heterostructure at the BSE level under different temperatures of 200 K, 300 K, and 400 K. (c) Oscillator strength of excitons in the energy range of 2 – 5 eV, considering electron-phonon coupling at 300 K. (d) Real-space wave function of the first bright exciton (second of all) of the heterostructure, considering electron-phonon coupling at 300 K.